\documentstyle[aps,preprint]{revtex}

\tightenlines

\begin{document}

\draft 

\widetext

\title{Excited States of Linear Polyenes}

\author{William Barford$^{1\ast}$, Robert J. Bursill$^2$ and Mikhail Yu
Lavrentiev$^{1\ast\ast}$}

\address{
$^1$Department of Physics and Astronomy, The University of Sheffield,
\\
Sheffield,  S3 7RH, United Kingdom.
\\
$^2$School of Physics, University of New South Wales, Sydney, NSW
2052,
Australia.
}

\maketitle

\mediumtext

\begin{abstract}
We present density matrix renormalisation group calculations of the
Pariser-Parr-Pople-Peierls
model of linear polyenes within the adiabatic approximation.
We calculate the vertical and relaxed transition energies, and relaxed
geometries for various excitations on long
chains. The triplet ($1^3B_u^+$) and even-parity singlet ($2^1A_g^+$)
states  have a
2-soliton and 4-soliton form, respectively, both with large relaxation
energies. The dipole-allowed ($1^1B_u^-$) state forms an
exciton-polaron and
has a very small relaxation energy. The relaxed energy of the
$2^1A_g^+$ state lies below that of the $1^1B_u^-$ state.
We observe an attraction  between the soliton-antisoliton
pairs in the $2^1A_g^+$
state.
The calculated excitation energies agree well with the
 observed values for polyene oligomers; the agreement with polyacetylene
 thin
 films is less good, and we comment on the possible sources of
 the discrepencies.
The photoinduced absorption is interpreted.  The
spin-spin
 correlation function shows that the unpaired spins coincide
with the geometrical soliton positions.  We study the roles
 of electron-electron
interactions and electron-lattice coupling in determining the excitation
energies and soliton structures.  The electronic
interactions play the key role in determining the ground state dimerisation
and the excited state transition energies.
\end{abstract}

\pacs{PACS numbers: 71.10.F, 71.20.R, 71.35}

 \narrowtext

\section{Introduction}

The inter-play of electron-electron interactions and
electron-lattice
coupling in linear polyenes
results in
a wealth of  low-lying excitations.
Electron-electron interactions
induce spin density wave correlations in the ground
state.  The lowest lying excitations are
 triplets, which combine to form  dipole-forbidden
singlet
($^1A_g^+$) excitations.   Optical excitations are
gapped, lie above the  $2^1A_g^+$ state, and are essentially
ionic
in
character, that is, there is charge transfer from one site to another.
The lowest optically allowed ($1^1B_u^-$) state  lies below
the charge gap \cite{footnoteI}, and is thus excitonic in character.  For
convenience, we show the
group
theoretic labelling of the states discussed in this paper in
Table I.

Electron-phonon interactions result in a
dimerised semiconducting ground state.
Within the adiabatic approximation, the
non-linear excitations include charged-spinless  and neutral-spin 1/2
solitons.  Both
electronic
interactions and electron-lattice coupling lead to
a gap in the optical spectrum.
In contrast to the interacting limit, however, the
$2^1A_g^+$ state always lies above the $1^1B_u^-$ state in the
non-interacting electron-phonon model.

The realisation that electronic interactions play a significant role in
polyenes
came via
the experimental observation,
by Hudson and Kohler\cite{hudson72} in 1972, that the $2^1A_g^+$ state lies
below the
$1^1B_u^-$
state.  At the same time, by
perfoming
a double configuration interaction calculation on the
Pariser-Parr-Pople
model,
Schulten and Karplus\cite{schulten72} demonstrated that the $2^1A_g^+$
wavefunction has a strong triplet-triplet contribution, and has a lower
energy
than the
$1^1B_u^-$ state.  The triplet-triplet and correlated
nature of the $2^1A_g^+$ state has
been further
investigated by Tavan and Schulten\cite{tavan87} and other workers\cite{refs}.
In 1986, Hayden and
Mele\cite{hayden86} performed a real space renormalisation group
calculation on the
Hubbard-Peierls model of up to sixteen sites and found that the $2^1A_g^+$
state was
composed of 4-solitons. This 4-soliton nature has
also been investigated by Su
\cite{su95},
and Wen and Su\cite{wen97}.
Ovchinnikov {\em et al.} also high-lighted the role of electronic
interactions,
by suggesting that they are largely responsible for the optical gap
\cite{ovchinnikov}.
In contrast to the strong deviations from the ground state geometry
predicted
for the
triplet and $2^1A_g^+$ state, Grabowski {\em et al.}\cite{grabowski85}
predicted
that the $1^1B_u^-$ state is an exciton-polaron.

The existence of the $2^1A_g^+$ state below the $1^1B_u^-$ state in
polyacetylene
thin films has been suggested by a number of experiments.  Third harmonic
generation
(THG) and two photon absorption by Halvorson and
co-workers\cite{halvorson93}
indicate that a $^1A_g^+$ state lies below 1.1 eV; while
the linear absorption,
locating
the
$1^1B_u^-$ state,  typically rises at 1.8 eV and peaks at
2.0
eV \cite{vardeny}.  However, Fann
and co-workers\cite{fann89} performed THG, finding peaks at 0.6 eV and
0.89 eV,
which they interpret as $^1A_g^+$  and $^1B_u^-$ states virtually
coincident at 1.8
eV.
The position of the $2^1A_g^+$ state is therefore
 not definitively established.
We return to this point in section V when we discuss our own theoretical
predictions.
For a detailed review
of the experimental and theoretical studies of conjugated polymers up to 1992,
see \cite{kiess}.

Electron-electron interactions in
$\pi$-conjugated systems, such as  {\em trans}-polyacetylene,
are conveniently
modelled
by the
one-band
Pariser-Parr-Pople model, which includes long range Coulomb interactions.
This semi-empirical model has been extensively used to study the excited
states of small conjugated molecules with a remarkable degree of success
\cite{bursill98}.
The Peierls model describes the electron-lattice coupling in the adiabatic
limit.
Thus, the Pariser-Parr-Pople-Peierls model is a realistic and accurate model
of $\pi$-conjugated systems, which captures their essential physics.
In an earlier paper\cite{bursill99} we performed accurate calculations
on this model using
 the infinite lattice
algorithm of
the
density matrix renormalisation group (DMRG)
method \cite{White}, \cite{book}.
The
Hellmann-Feynman
 theorem
was used to calculate the low-lying excited states and the lattice
geometry associated with them.   We showed that the $1^3B_u^+$ and
$2^1A_g^+$
states are modelled by 2 and 4 soliton fits, respectively, and that the
$1^1B_u^-$ state
is an exciton-polaron.  In this paper we develop that work.  In particular,
our objectives are:

\begin{enumerate}

\item
To further demonstrate that the DMRG calculations are reliable by, (i)
making
comparisons to the exact non-interacting limit, and (ii) comparing the
 infinite
lattice method to the finite lattice method.

\item
Use a realistic model of polyenes to understand the roles
 of electron-electron
interactions and electron-lattice coupling in determining the
dimerisation of the ground state and the transition energies of the excited
states.  In agreement with the earlier work of
Horsch\cite{horsch81}, and Konig and Stollhoff \cite{konig90}, we
find that  the electronic interactions play the key role in driving the ground
state
dimerisation.  Electronic interactions are also dominant
 in determining the solitonic structures and transition energies of the
excited states.

\item
To make more detailed comparisons to other experimental
 probes, in particular
photo-induced absorption.
The agreement with a wide range of experiments
confirms the validity of the model, our calculational method,
and our predictions on the soliton structures and their interactions.

\item
To further investigate both the geometry and electronic properties of solitons.

\end{enumerate}

This paper also serves as a correction to \cite{bursill99}.  In that paper
we used the dimerised ground state geometry in the Coulomb interactions
to calculate  the energy of all
the states.  Thus, the Coulomb interactions (unlike the one-electron transfer
integrals) were not updated in the Hellmann-Feynman minimisation procedure for
the relaxed states.  We find that using the correct geometry in the Coulomb
interactions affects the excitation energies by ca.\ $0.1$ eV.
The geometry of the triplet excited state is modified, so that now there is no
soliton-antisoliton confinement in the triplet state.  However, attractive
soliton interactions remain in the $2^1A_g^+$ state.

The plan of this paper is as follows.  In the next section we introduce
the Pariser-Parr-Pople-Peierls model.  To establish the consequences
of the inter-play of electron-electron interactions and
electron-lattice
coupling, we consider these two limits separately in sections
 III and IV.  The
non-interacting
limit also allows us to compare the infinite and finite DMRG algorithms
to an exact
calculation.
In section V we solve the full model, and discuss
the vertical and relaxed energies
of the key excited states. As well as
linear
absorption and non-linear optical spectroscopies, photo-induced
absorption is a useful tool in determining the positions of excited
states.
We discuss the experimental situation and our theoretical interpretation
 in section VI.   In section VII we consider the solitonic structures.  By
making comparisons  between the
geometrical soliton structures and the spin-spin correlation functions,
we show how they are closely related.
We conclude and discuss in section VIII.

As well as the work already mentioned, earlier work on the solitonic
structure of the
low-lying excitations include,  a mean-field study of the
Heisenberg-Peierls
model
\cite{takimoto89} and an exact diagonalisation of a 12 site extended
Hubbard-Peierls
model \cite{gammel93}. The DMRG method has recently been used by
Yaron et al.\ \cite{yaron98} and Fano et al.\ \cite{fano98} to solve the
Pariser-Parr-Pople
model for linear and cyclic polyenes, respectively.

\section{The Pariser-Parr-Pople-Peierls Model}

The Pariser-Parr-Pople-Peierls model is a realistic and accurate model
of $\pi$-conjugated systems, which includes the key features
of long range electron-electron interactions and electron-lattice
coupling.  The Hamiltonian for an $N$ site chain with open
boundary conditions is defined as
\begin{eqnarray}
{\cal H}
& &
= - 2\sum_{\ell=1}^{N-1} t_{\ell} \hat{T}_{\ell}
+ \frac{1}{4 \pi t_0 \lambda} \sum_{\ell=1}^{N-1} \Delta_{\ell}^2 +
\Gamma \sum_{\ell=1}^{N-1} \Delta_{\ell} \nonumber \\
& &
+\;
U \sum_{i=1}^N \left(n_{i\uparrow}- \frac{1} {2} \right)
\left(n_{i\downarrow} -  \frac{1} {2} \right)
+ \sum_{<ij>}  V_{ij} (n_i - 1)(n_j - 1),
\end{eqnarray}
where, $<ij>$ indicates all pairs of sites,
$t_{\ell} = \left( t_0 + \frac{\Delta_{\ell}}{2} \right) $ and
\begin{eqnarray}
\hat{T}_{\ell}
= \frac{1} {2} \sum_{\sigma}
(c_{\ell+1 \sigma}^{\dagger} c_{\ell \sigma} +  h.c.)
\end{eqnarray}
is the bond order operator of the $\ell$th bond. We use the Ohno function
for
the
Coulomb interaction:
\begin{eqnarray}
V_{ij} = U / \sqrt{ 1 + \beta r_{ij}^2 },
\end{eqnarray}
where
$\beta = (U/14.397)^2$
and bond lengths are in \AA.
The dimensionless
electron-phonon coupling constant,
$\lambda$, is defined by
\begin{eqnarray}
\lambda = \frac{2 \alpha^2}{\pi K t_0},
\end{eqnarray}
where $K$ is the elastic spring constant (estimated to be 46 eV
$\AA^{-2}$ from Raman analysis of C-C stretching modes in
{\em trans-}(CH)$_2$) \cite{ehrenfreund87}, and $\alpha$ relates
the actual distortion of the $\ell$th.\ bond from equilibrium,
$\delta r_{\ell}$,
to $\Delta_{\ell}$:
\begin{eqnarray}
\delta r_{\ell} = \Delta_{\ell} / 2 \alpha.
\end{eqnarray}
We take the undistorted chain to lie along the $x-$axis, with the
bonds oriented at $30^o$ to this axis.  Then, for fixed bond angles,
the distorted chain coordinates are defined as:
\begin{eqnarray}
x_{ij} = x^0_{ij} - \frac{\sqrt{3}}{4\alpha}
\sum_{\ell=i}^{j-1} \Delta_{\ell},
\nonumber
\end{eqnarray}
\begin{eqnarray}
y_{ij} = y^0_{ij} - \frac{1}{4\alpha} \sum_{\ell=i}^{j-1} \Delta_{\ell}
(-1)^{\ell +1 },
\end{eqnarray}
where
\begin{eqnarray}
x_{ij}^0 = \frac{\sqrt{3}}{2} a_0 |j-i|
\nonumber
\end{eqnarray}
and
\begin{eqnarray}
y_{ij}^0
& &
 = 0, {\rm if }\ |j-i|\ {\rm even} \\
\nonumber
& &
=\;
\frac{a_0}{2} (-1)^{(i+1)},\ {\rm otherwise}.
\nonumber
\end{eqnarray}
$a_0$ ($=1.40 \AA$) is the undistorted C-C bond length.

The force per bond, $f_{\ell}$ is
\begin{eqnarray}
f_{\ell} =
-\frac{ \partial \langle {\cal H} \rangle}
 {\partial \delta r_{\ell}}.
\end{eqnarray}
Using the Hellmann-Feynman theorem this can be re-written as,
\begin{eqnarray}
f_{\ell}
& &
= -2\alpha\left(
\frac {\Delta_{\ell}} {2\pi t_0\lambda} + \Gamma -
\langle T_{\ell} \rangle \right) \\
\nonumber
& &
-\;
\sum_{<ij>}' \frac{U \beta} {2\alpha(1+ \beta r_{ij}^2)^{3/2}}
\left( \frac{\sqrt{3}}{2} x_{ij}
+ \frac{ (-1)^{(\ell+1)}}{2} y_{ij}
\langle (n_i - 1)(n_j - 1) \rangle \right).
\end{eqnarray}
The prime over the sum indicates that the sum runs over all pairs of sites
which span the $\ell$th. bond.  The contribution to the bond force from the
Coulomb interaction is small compared to the kinetic energy term:  the
value of the Coulomb force from the nearest neighbor
density-density correlator
is approximately one tenth of the kinetic term.  Moreover, the
density-density correlator alternates in sign
and drops to less than one tenth of the nearest neighbor
density-density correlator,
 so the sum over all bonds
is also small. Table II shows the correlator for
up to five nearest neighbors.
We therefore only include the nearest neighbor density-density
correlator
in the evaluation of the distorted geometry.
(However, the full distorted geometry is used in the evaluation of the Coulomb
interaction, Eqn.\ (3).)

Using this approximation, and setting
$f_{\ell} = 0$, the self-consistent
equation for the equilibrium $\Delta_{\ell}$ is:
\begin{eqnarray}
\Delta_{\ell} = \left( \frac{ 2\pi\alpha t_0\lambda}
{\alpha -C_{\ell} t_0 \lambda} \right)
\left( \langle T_{\ell} \rangle -\Gamma - C_{\ell} a_0 \right),
\end{eqnarray}
where,
\begin{eqnarray}
C_{\ell} = \frac{U\beta} {2\alpha(1+ \beta(a_0 + \delta r_{\ell})^2)^{3/2}}
\langle (n_{\ell} - 1)(n_{\ell+1} - 1) \rangle.
\end{eqnarray}
We observe that, since the nearest neighbor density-density correlator
is negative, the Coulomb interactions tend to increase the bond
dimerisation.

The calculations were performed for fixed chain lengths, which is
enforced by setting,
\begin{eqnarray}
\Gamma = \frac{1}{N-1} \sum_{\ell = 1}^{N-1} \left( \langle T_{\ell}
\rangle - C_{\ell} a_0 \right).
\end{eqnarray}

To complete our discussion of the model we turn to its parametrisation.
There are three parameters in the model: $t_0$, $U$ and $\lambda$.  An
optimal parametrisation for $t_0$ and $U$ was found in \cite{bursill98}
by fitting the Pariser-Parr-Pople model to the excited states of benzene.
Assuming that this parametrisation is transferable between all
$\pi$-conjugated
systems, we use them here, and set $t_0 = 2.539$ eV and $U = 10.06$ eV.
The remaining parameter, $\lambda$, is found by fitting the
vertical energies of the $1^1B^-_u$ and $2^1A^+_g$ states,
calculated from the Parsier-Parr-Pople-Peierls model,
to the 6-site linear polyene \cite{bursill99}. This gives
$\lambda = 0.115$.  Finally, using $K = 46$ eV
$\AA^{-2}$  implies $\alpha = 4.593$ eV $\AA^{-1}$.

\section{Solution of the Peierls Model}

As originally recognised by Pople and Walmsley\cite{pople}, the
low lying excitations of the dimerised even $N$ site chain correspond to the
creation of two mid-gap states.  These excitations are associated
with localised geometrical structures which lead to a reversal
of the lattice dimerisation, and were subsequently
termed solitons.  The defect
states repel from each and are repelled from the ends of the chain.
Thus,
they reside
at approximately
$N/4$ and $3N/4$ along the chain (as may be seen in Fig.\ 5(b)).
Fig.\ 1 shows
a
schematic
energy
diagram of the molecular orbitals and defect states,
while Fig.\ 2 shows the energies of the $1^1B^-_u$ and $2^1A^+_g$ states
as a function of inverse chain length.  It is clear
that the first excited even parity state lies above the odd parity
state.  However, in the long chain, continuum limit, these states
are degenerate, with energy
$4\Delta_0/\pi = 0.12$ eV, using $\lambda = 0.115$,
$t_0 = 2.539$ eV and \cite{ssh}
\begin{eqnarray}
\Delta_0 = 8t_0 \exp\left[ - \left( 1 + \frac{1}{2\lambda} \right)
\right].
\end{eqnarray}
This gap is only a fraction of the experimentally measured gap of
approximately
2.0
eV\cite{footnoteIII}.  While a larger optical gap can be obtained
by increasing $\lambda$ and $t_0$, the energetic
ordering of the low lying
states would still be incorrect.  As we see in the next section, it is
electronic interactions which primarily open the optical gap, and reverse
the energetic ordering of the states.
Furthermore,  electronic interactions significantly modify
the soliton structures, as we show in section VII.

The non-interacting limit enables us to make a comparison between the
DMRG
methods and the
exact calculation.  In Fig.\ \ref{comparison1} the energy difference
between the
exact results and DMRG calculations is shown for the $1^1B^-_u$ and
$2^1A^+_g$
states. We see that
for both states the accuracies of the infinite and finite lattice
algorithm calculations are close, so that both methods can be
used in the actual calculations. The accuracy
is better for the $1^1B^-_u$ state, but even for the $2^1A^+_g$
the error is about 0.002 eV for the 50 site  chain in the
infinite lattice algorithm calculation. Other DMRG convergence
 tests, confirming
the validity of the method, were
presented
in \cite{bursill99}.

\section{Solution of the Pariser-Parr-Pople model}

The uniform chain in the limit of only on-site Coulomb interactions
is described by the Hubbard model.  At half-filling, the spin
excitations are gapless in the infinite chain limit, whereas the charge
excitations are gapped.
Even though the Pariser-Parr-Pople model contains long range interactions,
the spin excitations still appear to be gapless in the uniform chain,
 as shown in
Fig.\ 2.
The $2^1A^+_g$ state is also gapless, confirming the interpretation of it
as a pair of bound magnons.
The optical gap (E($1^1B_u^-$)) extrapolates to
approximately $1.6$ eV, and is
 excitonic, lying approximately $1.0$ eV below the
charge gap for long chains.  As discussed in section I, the energies of the
$2^1A_g^+$ and
$1^1B_u^-$ states
in polyacetylene thin films are believed to be at approximately $1.0
-1.8$, and
$2.0$ eV,
respectively.  Approximately $0.3$ eV should be deducted from the
calculated  $1^1B_u^-$ energy to account for solvation effects
\cite{yaron97}, indicating that the undimerised Pariser-Parr-Pople model
underestimates the optical gap by approximately $0.7$ eV and the
$2^1A_g^+$ energy by up to  $1.8$ eV.

\section{Solution of the Pariser-Parr-Pople-Peierls model}

Sections III and IV indicate that neither electron-lattice coupling nor
electron-electron interactions alone are sufficient to explain the
low energy excitations of polyene oligomers.  A pure electron-phonon
model predicts degenerate $1^1B_u^-$ and
$1^3B_u^+$ states with the $2^1A_g^+$ state lying above them, while a
pure electron interaction model underestimates the optical gap,
has gapless spin excitations
and does not lead to  a dimerised chain.  We now turn to the DMRG
solution of the Pariser-Parr-Pople-Peierls model.   We note that
an infinitesimally small electron-phonon coupling will open a gap in the
spin excitation spectrum for all electronic interaction strengths.

We first
calculate the ground state energy and lattice geometry.
The normalised staggered
bond dimerisation is defined as,
\begin{eqnarray}
\delta_{\ell} \equiv (-1)^{\ell} \frac {(t_{\ell} - \bar{t}) } { \bar{t}},
\end{eqnarray}
where $\bar{t}$ is the average value of $t_{\ell}$ in the middle of the
chain.   $\delta = 0.102$ in the center of the chain.  Using
$\alpha = 4.593$ eV $\AA^{-1}$, this implies that the
bond length alternation of the
ground state in the middle of the chain is $0.056$ \AA,
in close agreement with the experimental result of $0.052$ \AA
\cite{kahlert87}.

Using the ground state geometry, the vertical
energies (that is, the energies of these states with the ground state
geometry) ($E^{\text v}$) of the
$1^3B_u^+$,
$1^1B_u^-$ and
$2^1A_g^+$
states are calculated. These, as well as the relaxed energies
($E^{\text{0-0}}$), are shown
in Fig.\ 4(a) as a function of inverse chain
length.  The vertical energy of the $2^1A_g^+$
state lies approximately $0.3$ eV above that of the $1^1B_u^-$ state
in the long chain limit\cite{footnoteIV}.
The relaxation energy of the $1^1B_u^-$ state is modest, being
approximately
 0.2 eV
 for $102$ sites.
By contrast, the relaxation
energies of the $1^3B_u^+$ and $2^1A_g^+$ states are substantial,
being approximately $0.8$
eV and $1.5$ eV, respectively, and converge rapidly with increasing
chain length. The energy of the relaxed $2^1A_g^+$ state lies 1 eV below
that of the $1^1B_u^-$ state.
 We see in section VII that this strong relaxation is
associated with a large distortion of the ground state structure.

In Fig.\ 4(b) we plot the charge gap,
\begin{eqnarray}
E(N+1) + E(N-1) -2E(N),
\end{eqnarray}
and the energy of the $1^1B_u^-$ state.
In the long chain limit the charge gap
represents the energy of an uncorrelated electron-hole pair, and
therefore represents the band edge.  The relaxation energy of the
charge gap is roughly double that of the $1^1B_u^-$ state.  This
is to be expected, as the two charges form independent polarons, whereas
the excitonic $1^1B_u^-$ state forms a single polaron, as discussed
in section VII.
We see that the single chain binding energy is $2.4$ eV. However, the
unbound pair is strongly solvated (ca.\ $1.5$ eV), while the exciton
is more weakly solvated (ca.\ $0.3$ eV) \cite{yaron97}.
  This implies that the
bulk binding energy of the $1^1B_u^-$ state is ca.\ $1$ eV.

The experimental
values of $E^{\text{0-0}}(1^1B_u^-)$ and
$E^{\text{0-0}}(2^1A_g^+)$ for short
polyenes
are also
shown\cite{kohler}. The $2^1A_g^+$ values are in excellent agreement
with our calculation. The
$1^1B_u^-$ values are approximately
 $0.3$ eV lower than our predictions,
which is approximately the reduction expected by the solvation of the
chains in solution \cite{yaron97}.  Thus, for short polyene oligomers,
the optimised parametrisation of the
Pariser-Parr-Pople-Peierls model gives remarkably good results.

Kohler has analysed the experimental results for
$N = 6 -  16$\cite{kohler}.
For the $2^1A_g^+$ state the empirical relation,
\begin{eqnarray}
E^{\text{0-0}}(2^1A_g^+) = 0.96 + 20.72/N,
\end{eqnarray}
was derived.  This relation appears to confirm the work of
 \cite{halvorson93}, who
find a $2^1A_g^+$ state at $1.1$ eV in thin films.  However, there is no
particular reason why a linear extrapolation in $1/N$ is valid.
  Our calculation
for the Pariser-Parr-Pople-Peierls model shows a significant flattening
 off of the
$2^1A_g^+$ energy for chain lengths of roughly $30$ or more sites.  The
calculated converged energy of $1.74$ eV is in agreement with \cite{fann89}.

This rapid convergence of energy with chain length is in contrast to both
 the
Pariser-Parr-Pople and Peierls models.  In the Peierls model the excitation
 energies
are gapped, but the deviation from $1/N$ behavior is only evident for
 long chains
(ca.\ $100$ sites).  In the Pariser-Parr-Pople model a deviation from $1/N$
behavior is only evident in the long chain limit for the $1^1B_u^-$ state
 and the
charge gap.  In the Pariser-Parr-Pople-Peierls model, however, states
 which form
pronounced solitonic structures, such as the $2^1A_g^+$ and triplet
 states (as
discussed in section VII) self-trap once the chain length exceeds the
 size of their
solitonic structures.  It is possible that this self-trapping
 is a consequence of the
adiabatic treatment of the lattice, and that a full treatment
 involving quantum
phonons would change this prediction.

Our understanding of self-trapping - and its validity or otherwise - is
 complicated by
the discussion of the $1^1B_u^-$ state energy.  Again,
an empirical relation,
\begin{eqnarray}
E^{\text{0-0}}(1^1B_u^-) = 2.01 + 15.60/N,
\end{eqnarray}
was derived by Kohler, which is in good agreement with the
thin film result.  Our calculated value of $2.74$ eV is too high,
even when solvation effects (ca.\ $0.3$ eV) are deducted.  Once again, the
$1^1B_u^-$ state is self-trapped, and the possible relaxation by lattice
 fluctuations
would lead to a better agreement.

However, since the phonon frequency of ca.\ $0.2$ eV is so small compared
 to the
electronic energy scales, any corrections to the adiabatic limit
 are expected to be
small, so we need to consider other possible reasons for the
 discrepencies in the
long chain limit.  One source is the possible renomalisation of the
Pariser-Parr-Pople-Peierls model parameters in the long chain limit;
another is
$\sigma$-electron screening.

\section{Photo-induced absorption}

The photo-induced absorption spectrum of a
 system, obtained while it is being pumped
at an energy above the optical gap, gives an insight into the excited
states of that system \cite{vardeny}.    Typically the system is pumped
at $2.4$ eV, and photo-induced absorption peaks are observed at $0.43$ eV
and $1.35$ eV.

The higher energy peak is believed to intrinsic and has been ascribed
to a bound soliton-anti-soliton pair \cite{orenstein}.  A possible
interpretation is that excitations to states above the vertical
$1^1B_u^-$ state decay non-radiatively to the $2^1A_g^+$ state,
which subsequently relaxes.  The
photo-induced absorption is then a vertical transition from the relaxed
$2^1A_g^+$ state to a $^1B_u^-$ state.  We find
that the
energy of the $1^1B_u^-$ state in the relaxed geometry of the $2^1A_g^+$
state lies
1.3 eV above the $2^1A_g^+$ state for $102$ sites.  However, the dipole moment
is weak, being only $ 0.16 \langle \mu \rangle_{1^1B_u^-}$ (where
$\langle \mu \rangle_{1^1B_u^-}$ is the dipole moment between the ground state
and $1^1B_u^-$ state).  A second possibility is that it is a
triplet-triplet  ($T \rightarrow T^{\ast}$)
transition.  We calculate this transition energy
to be $2.8$ eV, while the dipole moment is
$0.96 \langle \mu \rangle_{1^1B_u^-}$.  Since the excited triplet
($T^{\ast}$) is
a triplet-exciton (as opposed to a spin-density wave excitation) at
high energy, it is reasonably to assume that it will be strongly solvated,
reducing this transition energy by as much as $1$ eV.
Thus, a triplet to triplet transition is a possible explanation
 for this absorption.

We calculate the transition energy between the lowest polaron state
and the first dipole connected excitation to be $0.45$ eV at $102$
sites, and the
dipole moment
is
$0.88 \langle \mu \rangle_{1^1B_u^-}$,
 suggesting that this is the origin of the lower peak.

\section{Soliton Structures}

In Fig.\ 5(a) we plot, as a function of bond index from the
center of the chain, the
normalised staggered bond dimerisation, $\delta_{\ell}$, Eqn.\ (14).
  We note that the $1^3B_u^+$ and
$2^1A_g^+$
states undergo considerable bond distortion, whereas
the $1^1B_u^-$ state shows a weak polaronic
distortion of the lattice, similar to the distortion associated with
 a doped charge.
In \cite{bursill99} we showed that the $1^3B_u^+$ and
$1^1B_u^-$ states
fit a 2-soliton form
\cite{su95}, \cite{campbell},   \cite{bach},
whereas the $2^1A_g^+$ state fits a 4-soliton form.
The bond distortions of the non-interacting limit (the Peierls model)
are plotted in Fig.\ 5(b).
A comparison between these plots
 illustrates the role played by the electronic
interactions
in
modifying the non-interacting picture:
\begin{enumerate}

\item
The dimerisation in the ground state is  enhanced by a five-fold factor, in
qualitative agreement with \cite{horsch81} and \cite{konig90}.

\item
The $1^1B_u^-$ state evolves to an exciton-polaron, in agreement with
\cite{grabowski85}.

\item
The $2^1A_g^+$ state, owing to its strong triplet-triplet contribution,
evolves to  a
4-soliton solution, in agreement with \cite{hayden86}.

\end{enumerate}

Further insight into the electronic structure of
polyenes and its relation to their geometry can be obtained from the
spin-spin correlation function, defined as,
\begin{equation}
S_i = -\langle S^Z_i S^Z_{N+1-i} \rangle.
\end{equation}
This function measures anti-ferromagnetic
correlations between sites symmetrically
situated with respect to the center of the chain.
As the correlation function shows unimportant
oscillations between even and odd site indices $i$,
we use  the symmetrized function:
\begin{equation}
\tilde{S}_j = \frac{1}{2} ( S_{\frac{1}{2}(N-j)}+S_{\frac{1}{2}(N-
j)+1}),
\end{equation}
$j=0,4,8,...,N-2$, which measures the correlations between
pairs of doubly-bonded sites, with $j$ being the distance between
them.

The spin-spin correlation
functions, calculated in the ground state
geometry, are shown in Fig.\ 6(a).
They show a monotonic decay
for the correlations in the $1^1A^+_g$ and $1^1B^-_u$ states,
but in the $2^1A^+_g$ state there is a small minimum at $j=8$
and a maximum at $j=16$.
This behavior of the spin-spin correlations in the $2^1A^+_g$ state
becomes  clearer when we calculate it in the relaxed geometry  for
this state. Here, the correlation function of the $2^1A^+_g$ state,
shown in Fig.\ 6(b), has a strong minimum at $j=8$,
where it changes sign, and a maximum at $j=20$. These features
strongly confirm the triplet-triplet character of this state.  By
comparing
Fig.\ 6(b) to
 the soliton structure shown in Fig.\ 5(a), we
see
that the
unpaired spins correspond to the positions of the geometrical solitons.

\section{Discussion and Conclusions}

We began this investigation of the electronic and geometrical structure
of  linear polyenes by performing seperate studies of the
$U=0$ Peierls model and the $\lambda=0$ Pariser-Parr-Pople model.
These studies show that these two limits predict quite different
low-lying excitations.  The Peierls model predicts mid-gap states
associated with geometrical defects.  The dipole forbidden
$2^1A_g^+$ state lies above the degenerate singlet and triplet
$1B_u$ states.  In contrast, the Pariser-Parr-Pople model predicts
gapless (or very small gapped) triplet and $2^1A_g^+$ states, with
the $1^1B_u^-$ state lying above them.

When these two models are combined in the Pariser-Parr-Pople-Peierls model
 we
see the effect of the inter-play of electron-electron and electron-phonon
 interactions.
The lowest lying triplet
($1^3B_u^+$) is a soliton-antisoliton pair; the lowest lying singlet
($2^1A_g^+$) is an even-parity pair of
soliton-antisoliton pairs, owing to it being a bound pair of triplets;
and the lowest optically allowed state ($1^1B_u^-$) is an
exciton-polaron.
The soliton positions in the $2^1A_g^+$ state
is confirmed by the spin-spin correlation function.
Electron-electron interactions play the dominant role in opening the
optical gap and dimerising the lattice.

We find that the relaxation energy of the $1^3B_u^+$ and $2^1A_g^+$
states are
substantial,
whereas that of the $1^1B_u^-$ state is modest.  The vertical energy of the
$2^1A_g^+$ state lies above  that of the $1^1B_u^-$ state, but the relaxed
$2^1A_g^+$ state lies ca.\ $1.0$ eV below that of the $1^1B_u^-$ state.
  The role
of electron-electron interactions are crucial and subtle in determining
 these relative
positions.  A larger electron-electron interaction leads to  a more
 dimerised ground
state, and this tends to raise the vertical energy of the $2^1A_g^+$ state
 relative to that
of the $1^1B_u^-$ state.  However, a larger electron-electron interaction
also leads to
a larger relaxation of the $2^1A_g^+$ state energy compared to that of the
$1^1B_u^-$ state, leading to a reversal of their energies.

For short polyenes we find good agreement with experimental values.
  However, in
the long chain limit the results (at least for the $1^1B_u^-$ state) become
 more
qualitative.  The experimental uncertainty in the position of the
 $2^1A_g^+$ state
means that we cannot be sure of the validity of our prediction.
  However, if we
assume that ca.\ $1.0$ eV is the correct relaxed energy of
 the $2^1A_g^+$ state,
then our predictions are between $0.5$ to $1.0$ eV too high.
  In section V we
discussed some of the possible origins of these discrepencies.
  They include, the
neglect of lattice fluctuations in the adiabatic treatment of the
 lattice, the possible
renormalisation of the $\pi$-model parameters in the long chain
 limit, and the neglect of
the $\sigma$-bond screening.
We would expect that as a molecule gets larger the $\pi$ orbitals
will become more extended, as they mix with other orbitals.  This will
reduce $U$ and $\alpha$ (and hence $\lambda$), and increase $t_0$, thus
reducing the excitation energies.
  Work is currently in progress to
 study these affects.

\vspace{2 cm}

{\bf Acknowledgments}

This work was started while W.\ B.\ was on Study Leave at the UNSW.  He
thanks
the
Royal Society and the Gordon Godfrey Committee of the UNSW for
financial
support.  R.\ J.\ B.\ was supported by the Australian Research Council
and
M.\ Yu.\ L.\ was supported by the EPSRC (U.K.) (GR/K86343).
We thank D.\ Yaron and Z.\ Vardeny for discussions.

\begin{table}[h]
\caption{
The classification of the relevant states.
}
\begin{tabular}{cccc}
State      &  $^1A_g^+$ & $^1B_u^-$ & $^3B_u^+$  \\
\hline
Spatial Inversion Symmetry &  + & $-$ & $-$     \\
Spin & 0 & 0 & 1    \\
Particle-hole symmetry     & + & $-$ & +     \\
Character &  Covalent & Ionic & Covalent \\
\end{tabular}
\label{states}
\end{table}

\begin{table}[h]
\caption{
The density-density correlator as a function of distance
}
\begin{tabular}{cc}
$j$     &  $\langle (n_i - 1)(n_{i+j} - 1) \rangle$   \\
\hline
 1 & $-0.308$   \\
 2 & $+0.002$    \\
 3  & $-0.021$     \\
4 &  $+0.004$ \\
5 &  $-0.011$ \\
\end{tabular}
\end{table}

\begin{figure}[p]
\caption{
Energy level diagram for the key low-lying states in the non-interacting
limit.
}
\label{MO}
\end{figure}

\begin{figure}[p]
\caption{
Transition energies for the $1^1B_u^-$  (squares),  $2^1A_g^+$
(diamonds),
$1^3B_u^+$ (triangles) states and charge gap (circles)
as a function of inverse chain length for the $U=0$ Peierls model
(dashed lines and open symbols) and the $\lambda=0$
Pariser-Parr-Pople model (solid lines and filled symbols).
(In the Peierls model the $1^1B_u^-$ and $1^3B_u^+$ states are degenerate.)
}
\label{PPP}
\end{figure}

\begin{figure}[p]
\caption{
The difference between the exact calculation of the $2^1A^+_g$
(diamonds)
and $1^1B^-_u$ (squares) states in the non-interacting limit, and the
results
of DMRG calculations in the infinite and finite algorithms.
Solid lines correspond to infinite lattice algorithm results,
dashed lines to the finite lattice algorithm.
}
\label{comparison1}
\end{figure}

\begin{figure}[p]
\caption{
(a) Transition energies for the $1^1B_u^-$  (squares), $2^1A_g^+$
(diamonds)
and
$1^3B_u^+$  (triangles) states as a function of inverse chain length.
Vertical/relaxed transitions are indicated by dashed/solid lines
 and open/solid
symbols.
Experimental results for the relaxed $1^1B_u^-$ ($\times$) and
$2^1A_g^+$ ($+$)
state energies for polyenes in hydrocarbon solution
\protect\cite{kohler}.
(b) Transition energies for the $1^1B_u^-$  state (squares) and charge
gap (circles) as a function of inverse chain length.
}
\label{energies}
\end{figure}

\begin{figure}[p]
\caption{
(a) The geometries (normalised
staggered bond distortion $\delta_{\ell}$
 as a function of bond index $\ell$ from the
center of the lattice) of various states of the
Pariser-Parr-Pople-Peierls model: $1^1A_g^+$ (crosses),
$1^1B_u^-$
(squares),
$1^3B_u^+$ (triangles) $2^1A_g^+$ (diamonds) and polaron (circles),
for the $102$ site system.
(b) The same as (a) for the $U=0$ Peierls model.
}
\label{geometries}
\end{figure}

\begin{figure}[p]
\caption{
Spin-spin correlation functions for $1^1A^+_g$ (solid squares),
$2^1A^+_g$ (solid diamonds) and $1^1B^-_u$ (empty squares) states.
(a) In the relaxed $1^1A^+_g$ geometry,
(b) in the relaxed $2^1A^+_g$ geometry.
}
\label{correlations1}
\end{figure}


\begin{references}

\bibitem[*]{email}
Email address: W.Barford@sheffield.ac.uk.

\bibitem[**]{leave}
Current address: The School of Chemistry, University of Bristol,
Bristol, U.\ K.\
On leave from Institute of Inorganic Chemistry, 630090 Novosibirsk,
Russia.

\bibitem{footnoteI} As defined by the energy to add a particle and a hole
to the  half filled system.

\bibitem{hudson72} B.\ S.\ Hudson and B.\ E.\ Kohler, Chem.\ Phys.\
Lett.\
{\bf 14},
299 (1972).

\bibitem{schulten72} K.\ Schulten and M.\ Karpus, Chem.\ Phys.\ Lett.\
{\bf 14}, 305
(1972).

\bibitem{tavan87} P.\ Tavan and K.\ Schulten, Phys.\ Rev.\ B {\bf 36},
4337 (1987).

\bibitem{refs} R. R. Birge, K. Schulten and M. Karplus, Chem. Phys. Letts.
{\bf 31}, 451 (1975); R. McDiarmid, J. Chem. Phys. {\bf 79}, 1565 (1983);
B. Hudson and B. Kohler, Synth. Met. {\bf 9}, 241 (1984).

\bibitem{hayden86}
G.\ W.\ Hayden and E.\ J.\ Mele, Phys.\ Rev.\ B {\bf 34}, 5484 (1986).

\bibitem{su95}
W.\ P.\ Su,  Phys.\ Rev.\ Lett.\ {\bf 74}, 1167 (1995).

\bibitem{wen97} G-Z Wen and W-P Su,
Relaxations of Excited States and Photo-induced Structural Phase
Transitions (p 121),
Springer-Verlag 1997; Synth.\ Metals {\bf 78}, 195 (1996).

\bibitem{ovchinnikov} A.\ A.\ Ovchinnikov, I. I. Ukrainskii and
G.\ V.\ Kventsel, Sov.\ Phys.\ ---Usp.\ {\bf 15}, 575 (1973).

\bibitem{grabowski85} M.\ Grabowski, D.\  Hone and J.\ R.\ Schrieffer,
Phys.\ Rev.\
B {\bf 31}, 7850 (1985)

\bibitem{halvorson93} C.\ Halvorson and A.\ J.\ Heeger, Chem.\ Phys.\
Lett.\ {\bf
216}, 488 (1993).

\bibitem{vardeny} Relaxation in Polymers (p 174), ed. by T. Kobayashi.
World
Scientific (Singapore) 1993.

\bibitem{fann89} W.-S. Fann, {\em et al.}, Phys.\ Rev.\ Lett.\,
{\bf 62}, 1492 (1989).

\bibitem{kiess}  D.\ Baeriswyl, D.\ K.\ Campbell and S.\ Mazumdar, in
{\em Conjugated Conducting Polymers}, edited by H.\ Kiess (Springer-Verlag,
Berlin, 1992).

\bibitem{bursill98}
R.\ J.\ Bursill, C.\ Castleton, W.\ Barford, Chem.\ Phys.\ Lett.\ {\bf
294},
305 (1998).

\bibitem{bursill99} R.\ J.\ Bursill and W.\ Barford,
Phys.\ Rev.\ Lett.\,
{\bf 82}, 1514 (1999).

\bibitem{White}
S.\ R.\ White, Phys.\ Rev.\ Lett.\ {\bf 69}, 2863 (1992);
Phys.\ Rev.\ B {\bf 48}, 10345 (1993).

\bibitem{book} Density Matrix Renormalisation, edited by I.\ Peschel, X.\
Wang, M.\
Kaulke and K.\ Hallberg, Springer, Berlin, 1999.

\bibitem{horsch81} P. Horsch, Phys.\ Rev.\ B
{\bf 24}, 7351 (1981).

\bibitem{konig90} G.\ Konig and G.\ Stollhoff,
Phys.\ Rev.\ Lett.\,
{\bf 65}, 1239 (1990).


\bibitem{takimoto89}
J.\ Takimoto and M.\ Sasai, Phys.\ Rev.\ B {\bf 39}, 8511 (1989).

\bibitem{gammel93}
J.\ T.\ Gammel and D.\ K.\ Campbell, Synth.\ Met.\ {\bf 55}, 4638 (1993).


\bibitem{yaron98}
D.\ Yaron, E.\ E.\ Moore, Z.\ Shuai, J.\ J.\ Bredas, J.\ Chem.\ Phys.\
{\bf 108}, 7451 (1998).

\bibitem{fano98}
G.\ Fano, F.\ Ortolani, L.\ Ziosi, J.\ Chem.\ Phys\ {\bf 108}, 9246
(1998).

\bibitem{ehrenfreund87}
E.\ Ehrenfreund, Z.\ Vardeny, O.\ Barfman, B.\ Horovitz, Phys.\ Rev.\ B
{\bf 36},
1535 (1987).


\bibitem{pople} J.\ A.\ Pople and S.\ H.\ Walmsley, Molec.\ Phys.\  {\bf
5},
15
(1962).

\bibitem{ssh} A.\ J.\ Heeger, S.\ Kivelson, J.\ R.\ Schrieffer and
W-P
Su,
Rev.\ Mod.\ Phys.\ {\bf 60}, 781 (1988).

\bibitem{footnoteIII} However, the exact  $1^1B^-_u$ transition energy
obtained from Eq.\ (1) at 102 sites is
$0.23$ eV, while the $2^1A^+_g$ state is at $0.33$ eV.


\bibitem{yaron97}
E. Moore, B. Gherman B. and D. Yaron, J. Chem.\ Phys.\ {\bf 106}, 4216
(1997).

\bibitem{kahlert87} H.\ Kahlert, O.\ Leitner and G.\ Leising, Synth.
Met. {\bf 17}, 467 (1987).

\bibitem{footnoteIV}  Experimentally, this implies that the vertical energy
of the $2^1A^+_g$ state lies ca.\ $0.6$ eV above the $1^1B^-_u$ state,
because of the greater solvation
energy of the latter state.

\bibitem{kohler} B.\ E.\ Kohler, J.\ Chem.\ Phys.\ {\bf 88}, 2788 (1988).

\bibitem{orenstein} J.\ Orenstein and G.\ L.\ Baker, Phys.\ Rev.\ Lett.\,
{\bf 49}, 1043 (1982).

\bibitem{campbell}
D. K. Campbell and A. R. Bishop, Nucl. Phys. B {\bf 200}, 297 (1982).

\bibitem{bach} M. A. Garcia-Bach, R. Valenti, S. A. Alexander and D. J.
Klein,
Croatia Chemica Acta {\bf 64}, 415 (1991).


\end{references}
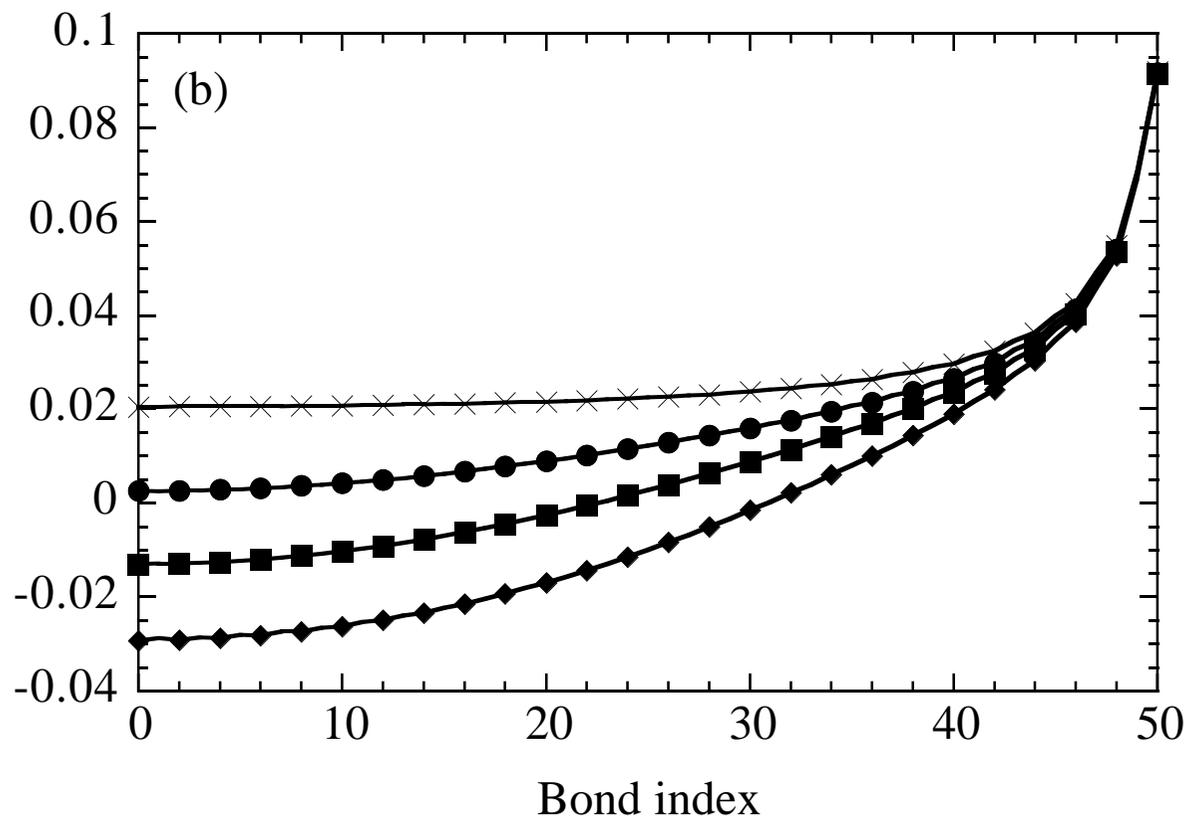
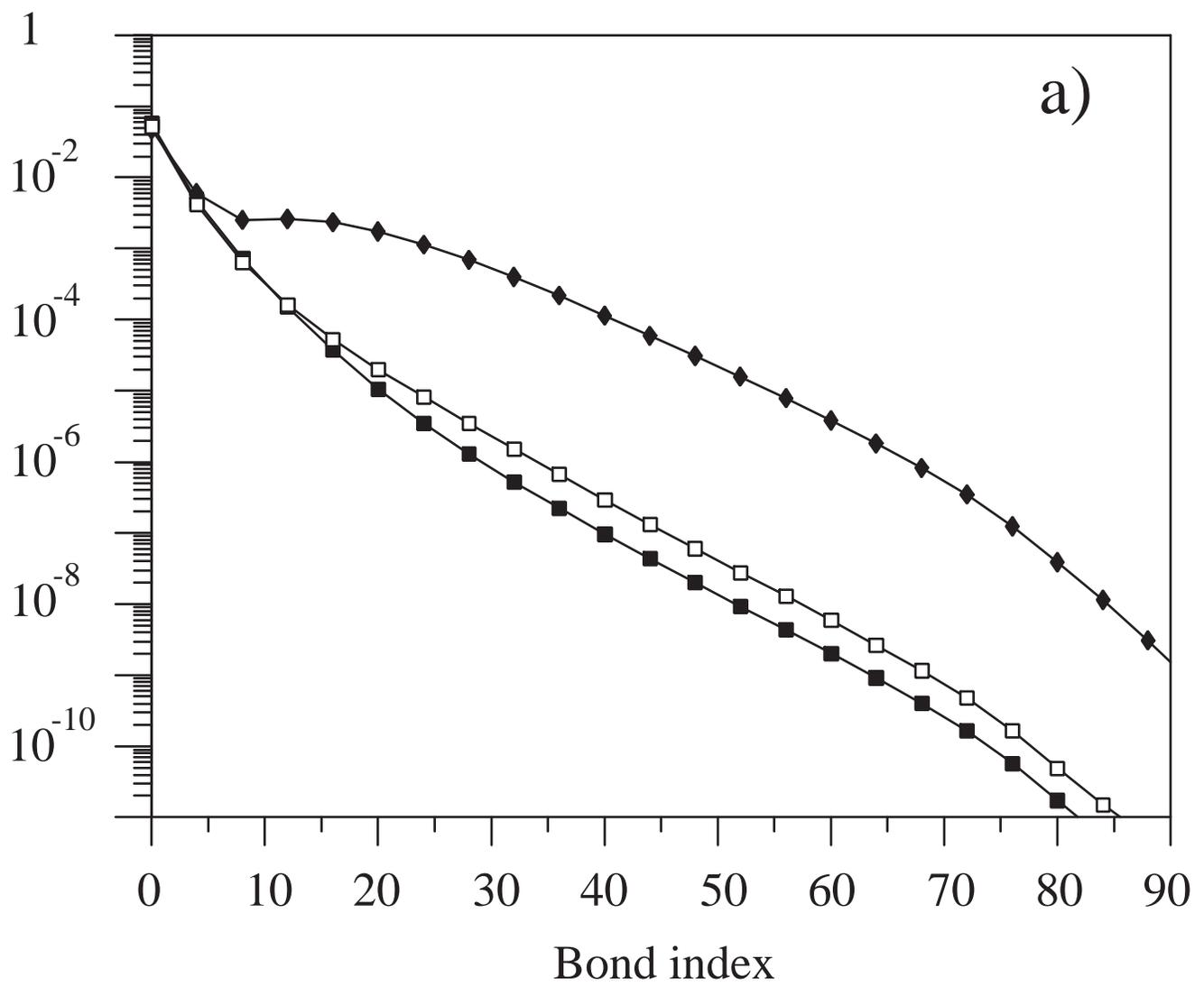
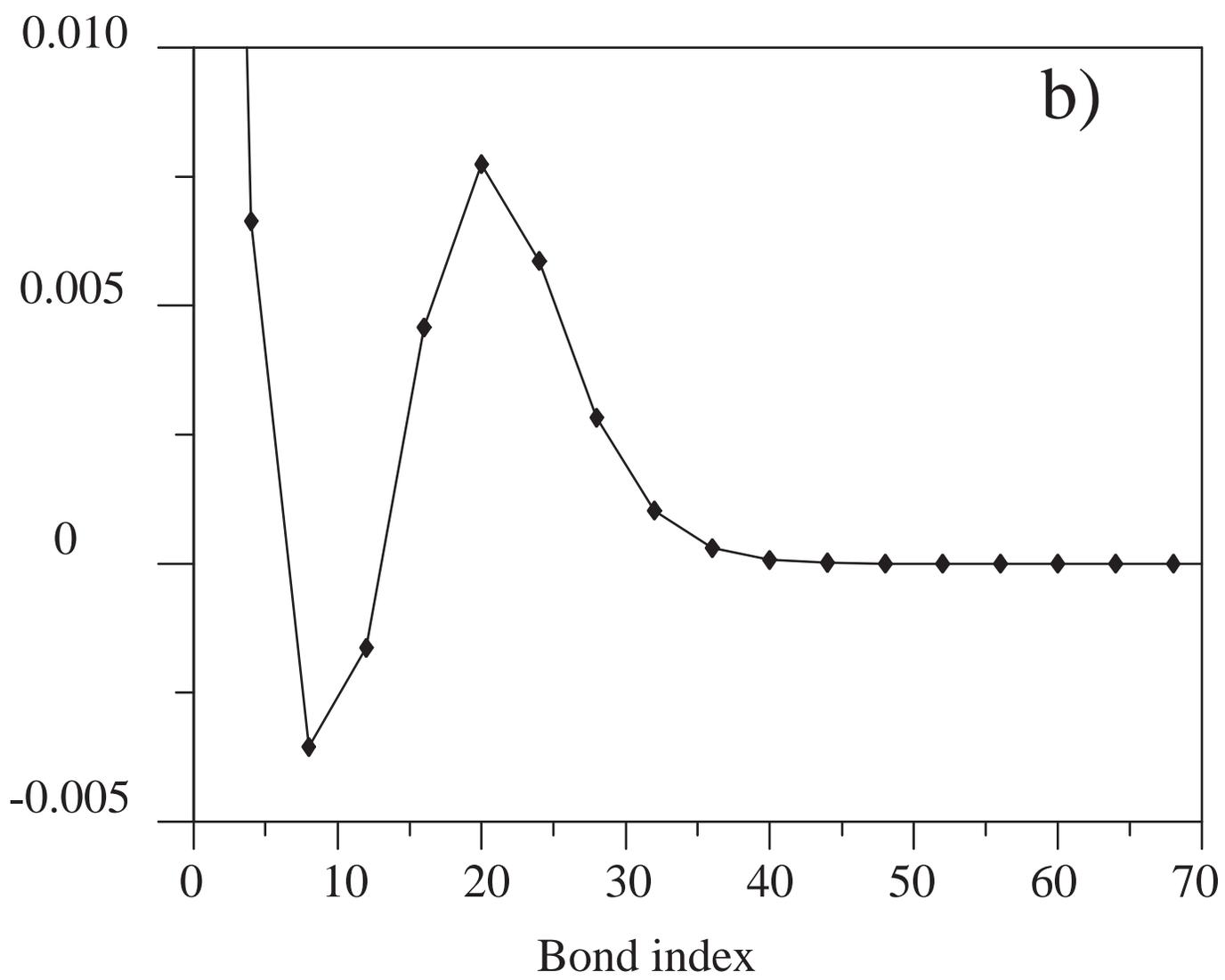
\end{document}